\documentclass[journal]{IEEEtran}
\ifCLASSINFOpdf
\else
\fi

\usepackage{cite}
\usepackage{graphicx}
\usepackage{subfig}
\usepackage{multirow}
\usepackage{float}

\begin{document}
%
\title{Global Ultrasound Elastography Using Convolutional Neural Network}
%
%
%

\author{Md. Golam Kibria$^1$,
        Hassan Rivaz$^{1,2}$\\
        $^1$Concordia University, Montreal, QC, Canada\\
        $^2$PERFORM Centre, Montreal, QC, Canada}

\maketitle

\begin{abstract}
Displacement estimation is very important in ultrasound elastography and failing to estimate displacement correctly results in failure in generating strain images. As conventional ultrasound elastography techniques suffer from decorrelation noise, they are prone to fail in estimating displacement between echo signals obtained during tissue distortions. This study proposes a novel elastography technique which addresses the decorrelation in estimating displacement field. We call our method GLUENet (GLobal Ultrasound Elastography Network) which uses deep Convolutional Neural Network (CNN) to get a coarse time-delay estimation between two ultrasound images. This displacement is later used for formulating a nonlinear cost function which incorporates similarity of RF data intensity and prior information of estimated displacement\cite{hashemi2017global}. By optimizing this cost function, we calculate the finer displacement by exploiting all the information of all the samples of RF data simultaneously. The Contrast to Noise Ratio (CNR) and Signal to Noise Ratio (SNR) of the strain images from our technique is very much close to that of strain images from GLUE. While most elastography algorithms are sensitive to parameter tuning, our robust algorithm is substantially less sensitive to parameter tuning.
\end{abstract}


%
\IEEEpeerreviewmaketitle

\section{Introduction}
%
%
%
%
Ultrasound is a non-invasive medical imaging modality which produces informative representation of human tissues and organs in real-time. Tissue deformation can be stimulated and imaged at the same time by manual palpation of the tissue using ultrasound probe. Estimation of tissue deformation is very important for ultrasound elastography. Elastography, a term proposed by Ophir et al.\cite{ophir1991elastography}, refers to a quantitative method for imaging elasticity of biological tissues. Ultrasound elastography can provide physicians with valuable diagnostic information for detection and/or characterization of tumors in different organs\cite{varghese2009quasi}.

Over the last two decades, many techniques have been reported for estimating tissue deformation using ultrasound. The most obvious approach is window-based methods with cross-correlation matching techniques. Some reported these techniques in temporal domain while others reported in spectral domain. Another notable approach for estimating tissue deformation is usage of dynamic programming with regularization and analytic minimization on one-dimensional (along axial) and two-dimensional (along axial and lateral) directions. All these approaches suffer severely from decorrelation noise and make a trade-off between image resolution and computational cost.

Tissue deformation estimation in ultrasound images is an analogous to the optical flow estimation problem in computer vision. The structure and elastic property of tissue imposes the fact that tissue deformation must contain some degree of continuity. So, tissue deformation estimation can be considered as a special case of optical flow estimation which is not bound by structural continuity. Apart from many state-of-the-art conventional approaches for optical flow estimation, very recently notable success has been reported at using deep learning network for end-to-end optical flow estimation. Deep learning networks enjoy the benefit of very fast calculation by trained (finetuned) weights of the network while having a trade-off of long-time computationally exhaustive training phase. A promising recent network called FlowNet 2.0\cite{ilg2017flownet} has achieved up to 140 fps at optical flow estimation. These facts indicate the promising potential for using deep learning approach for tissue deformation estimation using ultrasound images.

This work takes advantage of the fast FlowNet 2.0 architecture to estimate an initial time delay estimation which is robust from decorrelation noise. This initial estimation is then finetuned by optimizing a global cost function\cite{hashemi2017global}. This approach has many advantages over conventional methods. The most important one would be the robustness of the method to decorrelation noise of ultrasound images.

\section{Method}
The proposed method calculates the time delay between two radio-frequency (RF) ultrasound scans which are correlated by a displacement field in two phases combining fast and robust convolutional neural network with the more accurate global optimization based coarse to fine displacement estimation. This combination is possible due the fact that the global optimization-based method depends on coarse but robust displacement estimation which CNN can provide readily more robustly than any other state-of-the-art elastography methods.

\begin{figure*}
\centering
\fbox{{\includegraphics[width=\textwidth]{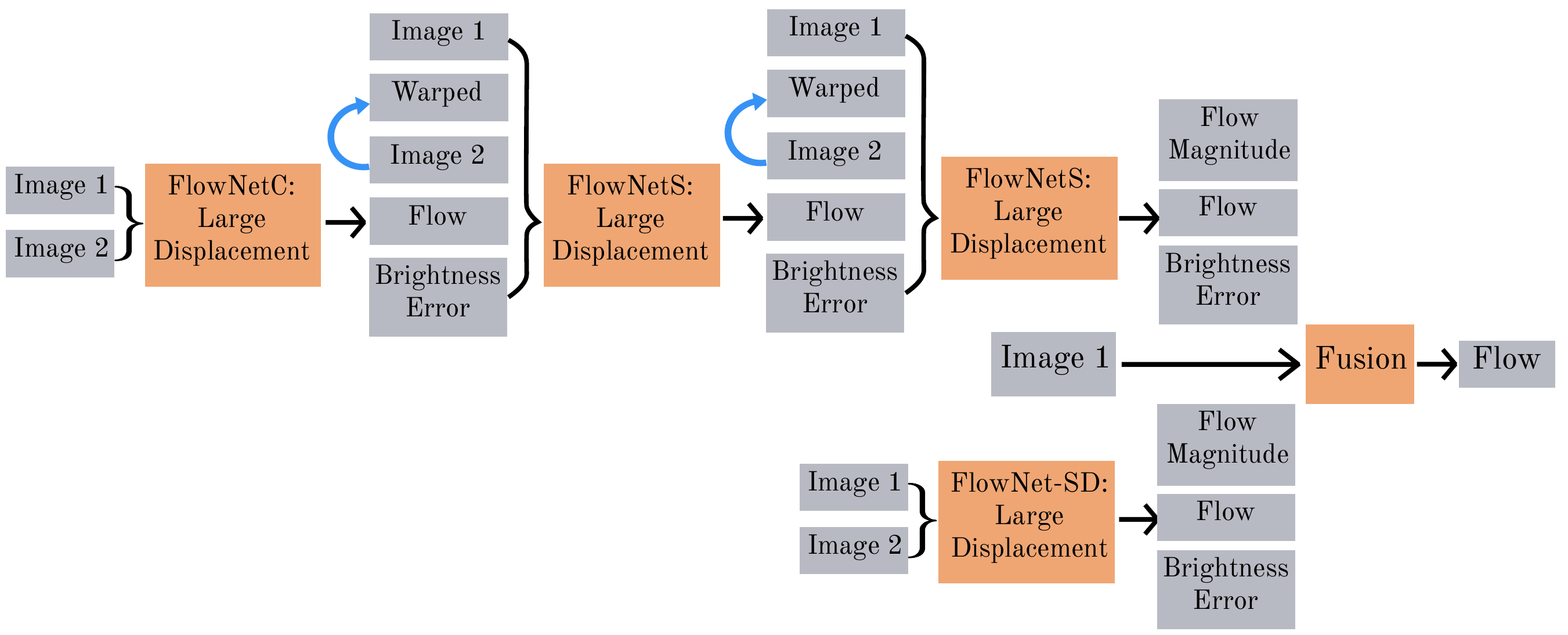}}}
\caption{ Full schematic of FlowNet 2.0 architecture. Braces indicate the concatenation of the inputs.} 
\label{fig:figFlownet2}
\end{figure*}

\begin{figure*}
\centering
\subfloat[]{\includegraphics[width=.33\textwidth]{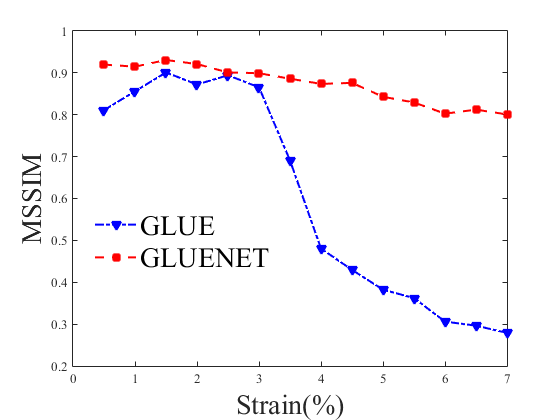}}
\subfloat[]{\includegraphics[width=.33\textwidth]{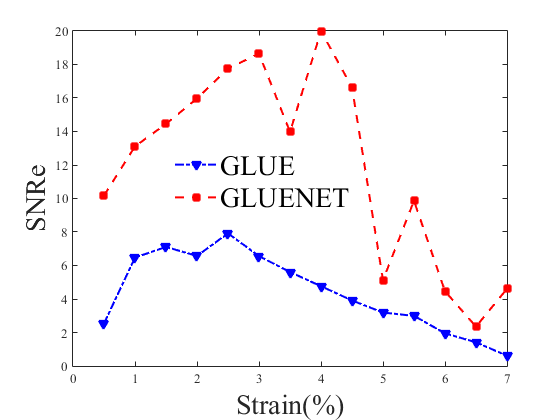}}
\subfloat[]{\includegraphics[width=.33\textwidth]{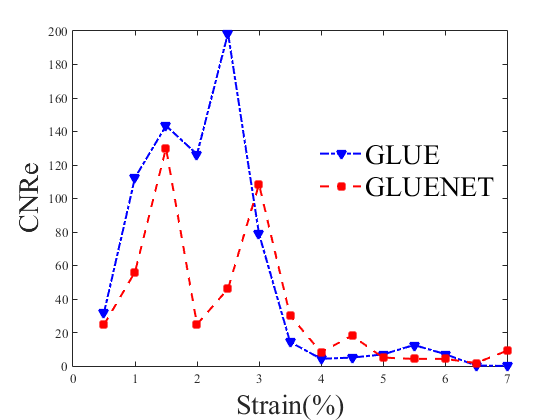}}
\caption{Performance comparison using simulation phantom data with added noise (PSNR 12.7dB) upto 7\% applied strain. (a) MSSIM (b) SNRe and (c) CNRe vs applied strain} 
\label{fig:matind}
\end{figure*}

Optical flow estimation in computer vision and tissue displacement estimation in ultrasound elastography share common challenges. So, optical flow estimation techniques can be used for tissue displacement estimation for ultrasound elastography. The latest CNN that can estimate optical flow with competitive accuracy with the state-of-the-art conventional methods is called FlowNet 2.0\cite{ilg2017flownet}. This network is an improved version of its predecessor FlowNet\cite{dosovitskiy2015flownet}, where Dosovitskiy et al. trained two basic networks namely FlowNetS and FlowNetC for optical flow prediction. FlowNetC is a customized network for optical flow estimation whereas FlowNetS is rather a generic network. The details of these networks can be found in\cite{dosovitskiy2015flownet}. These networks were further improved for more accuracy in\cite{ilg2017flownet} which is known as FlowNet 2.0.

Figure~\ref{fig:figFlownet2} illustrates the complete schematic of FlowNet 2.0 architecture. It can be considered as a combination of stacked version of FlowNetC and FlowNetS. This helps the network to calculate large displacement optical flow. Brightness error is the residual between the first image and the second image warped with the already estimated flow. For dealing with the small displacements small strides were introduced in the beginning. Also, convolution layers were introduced between upconvolutions in the FlowNetS architecture. Finally, a small fusion network estimates the final flow.

The displacement estimation from FlowNet 2.0 is robust but needs more refinement in order to produce strain images of high quality. Global Time-Delay Estimation (GLUE)\cite{hashemi2017global} is an accurate displacement estimation method provided that an initial coarse displacement estimation is available. If the initial displacement estimation contains large errors, then GLUE may fail to produce accurate fine displacement estimation. GLUE refines the initial displacement estimation by optimizing a cost function incorporating both amplitude similarity and displacement continuity. It is noteworthy that the cost function is formulated for the entire image unlike its motivational previous work\cite{rivaz2011am2d} where only a single RF line is optimized. The details of the cost function and its optimization can be found in\cite{hashemi2017global}. After displacement refinement, strain image is obtained by using least square or gradient based methods.

\begin{figure*}
\centering
\subfloat[GLUE]{\includegraphics[width=.49\textwidth]{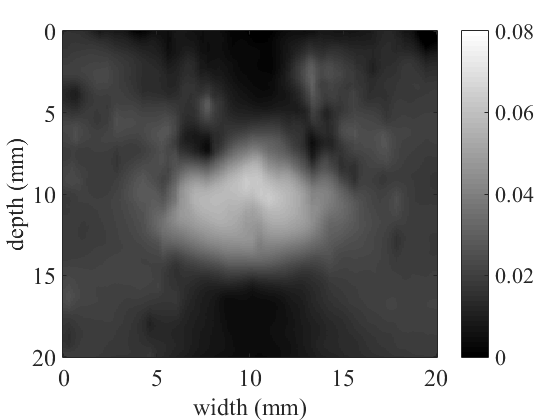}}
\subfloat[GLUENet]{\includegraphics[width=.49\textwidth]{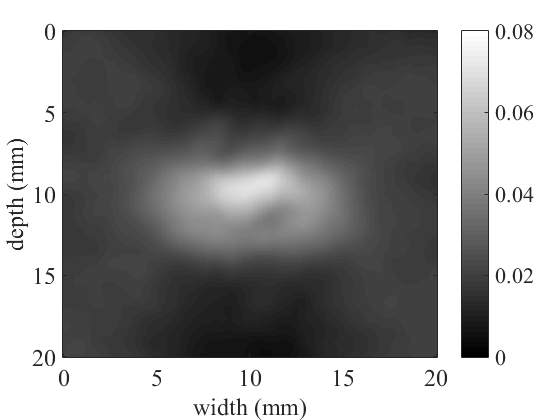}}
\caption{Strain images of simulation phantom with added noise (PSNR 12.7 dB)} 
\label{fig:sim}
\end{figure*}

\begin{figure*}
\centering
\subfloat[GLUE]{\includegraphics[width=.49\textwidth]{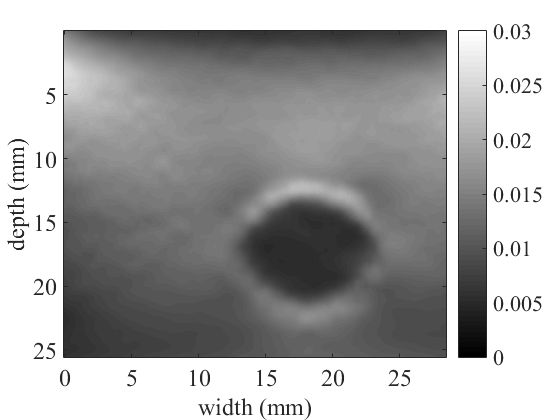}}
\subfloat[GLUENet]{\includegraphics[width=.49\textwidth]{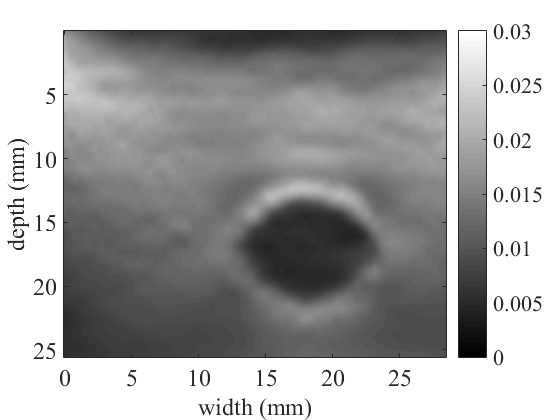}}
\caption{Strain images of CIRS breast phantom (a) GLUE (CNRe: 133.44, SNRe: 13.29), (b) GLUENet (CNRe: 132.2, SNRe: 13.11)} 
\label{fig:tmp}
\end{figure*}

\section{Results}
In this section, we present results of simulation and phantom experiments. The simulation phantom has a soft inclusion in the middle and the corresponding displacement is calculated using Finite Element Method (FEM) by ABAQUS Software (Providence, RI). The CIRS breast phantom (Norfolk, VA) has a single hard inclusion in the middle. RF data is acquired using an Antares Siemens system (Issaquah, WA) at the center frequency of 6.67 MHz with a VF10-5 linear array at a sampling rate of 40 MHz. Details of the data acquisition are available in\cite{rivaz2011am2d}.

For comparison of the robustness of our method we use mathematical metrics such as Mean Structural Similarity (MSSIM), Signal to Noise Ratio (SNR) and Contrast to Noise Ratio (CNR). Figure~\ref{fig:matind} illustrates the performance of the proposed method against GLUE\cite{hashemi2017global} in terms of these numerical metrics for simulation phantom with added noise (PSNR: 12.7 dB). Figure~\ref{fig:sim} demonstrates robustness of the proposed method to decorrelation noise using strain images. For demonstrating the effectiveness of our proposed method in CIRS phantom, we test our technique with 62 pre- and post-compression RF signal pairs from 20 RF signals. While GLUE fails to generate any recognizable strain images in 27 cases, our technique generates quality strain images for all 62 pairs proving the robustness of the technique. Figure~\ref{fig:tmp} shows the strain images of the CIRS phantom.

\section{Conclusion}
In this paper, we introduced a novel technique to calculate tissue displacement from ultrasound images using CNN. This is, to the best of our knowledge, the first use of CNN for estimation of displacement in ultrasound elastography. The displacement estimation obtained from CNN was further refined using GLUE\cite{hashemi2017global}, and therefore, we referred to our method as GLUENet. We showed that GLUENet is robust to decorrelation noise, which makes it a good candidate for clinical use.

\section*{Acknowledgment}
This research has been supported in part by NSERC Discovery Grant (RGPIN-2015-04136). We would like to thank Microsoft Azure Research for a cloud computing grant and NVIDIA for GPU donation. The ultrasound data was collected at Johns Hopkins Hospital. The principal investigators were Drs. E. Boctor, M. Choti, and G. Hager. We thank them for sharing the data with us.

\ifCLASSOPTIONcaptionsoff
  \newpage
\fi



\bibliographystyle{IEEEtran}
\bibliography{./bib/paper_db}
\end{document}